\pdfoutput=1
\documentclass[final,5p,times,twocolumn, a4paper]{elsarticle}

\usepackage{atbegshi}
\AtBeginShipout{%
  \ifnum\value{page}=1
    \setbox\AtBeginShipoutBox=\hbox{\hspace*{2mm}\box\AtBeginShipoutBox}%
  \fi
}
\usepackage{amssymb, amsmath, float, bm, graphicx, subcaption, lineno, mathtools}
\usepackage[numbers]{natbib}  

\graphicspath{{./plots/images}}

\journal{Journal of Computational Physics}

\begin{document}

\begin{frontmatter}

\title{Time-Domain Benchmark Solutions for Bernstein Waves}

\author[ipp,tum]{M. Pelkner\corref{cor1}} 
\author[ipp]{K. Hallatschek} 

\cortext[cor1]{Corresponding author. E-mail: maximilian.pelkner@ipp.mpg.de}

\affiliation[ipp]{organization={Max Planck Institute for Plasma Physics},
            addressline={Boltzmannstraße 2}, 
            city={Garching},
            postcode={85748}, 
            country={Germany}}

\affiliation[tum]{organization={TUM School of Natural Sciences},
            addressline={Boltzmannstraße 10}, 
            city={Garching},
            postcode={85748}, 
            country={Germany}}

\begin{abstract}

In previous work, we introduced a semi-analytical method for computing
time-domain solutions of linearized Vlasov problems. Rather than representing
the plasma response as a sum of residues associated with Landau poles,
the method constructs a regularized frequency-domain response spectrum that is
subsequently inverted numerically. Explicit applications have so far been
limited to unmagnetized plasmas. In this work, we extend the construction to
plasmas with a uniform background magnetic field and a Maxwellian equilibrium
distribution. We apply the resulting formulation to the electrostatic
ion-Bernstein density response of a plasma with kinetic ions and adiabatic
electrons, and provide error estimates for truncating both the spectral
integration domain and the cyclotron-harmonic expansion. The solution serves as
a time-domain reference for damped finite-$k_\parallel$ regimes, in which
dispersion-relation roots alone are insufficient for pointwise-in-time
verification of simulation codes. Finally, we indicate how the framework can be
extended to fully electromagnetic problems.

\end{abstract}

\end{frontmatter}

\section{Introduction}

To establish the reliability of simulations in analytically intractable
regimes, numerical codes must be carefully verified against analytical
benchmark solutions. In nuclear fusion research, this is particularly important
for kinetic plasma codes such as BSL6D~\cite{schild2024} and
GEMPIC~\cite{gempic}, which aim to resolve phase-space dynamics beyond the
standard gyrokinetic ordering.  

In previous work, we have developed a semi-analytical method that, in principle, allows to calculate time-domain solutions of linearized Vlasov problems with translationally invariant backgrounds~\cite{isw}. Rather than characterizing the asymptotic response in terms of complex-frequency
roots of the dispersion relation, the method constructs a rapidly decaying response spectrum and evaluates the inverse Fourier transform numerically. To reduce the algebraic complexity, explicit examples considered so far have been limited to unmagnetized plasmas or to wave propagation parallel to an external guide magnetic field~\cite{isw, cfsa}. Magnetized plasma dynamics, however, is strongly shaped by particle motion perpendicular to the background magnetic field~\cite{wesson}. Meaningful verification of kinetic codes therefore requires benchmarks with finite $k_\perp$.

\begin{figure}[!t]
    \centering
    \begin{subfigure}[t]{0.75\columnwidth}
        \centering
        \includegraphics[width=\linewidth]{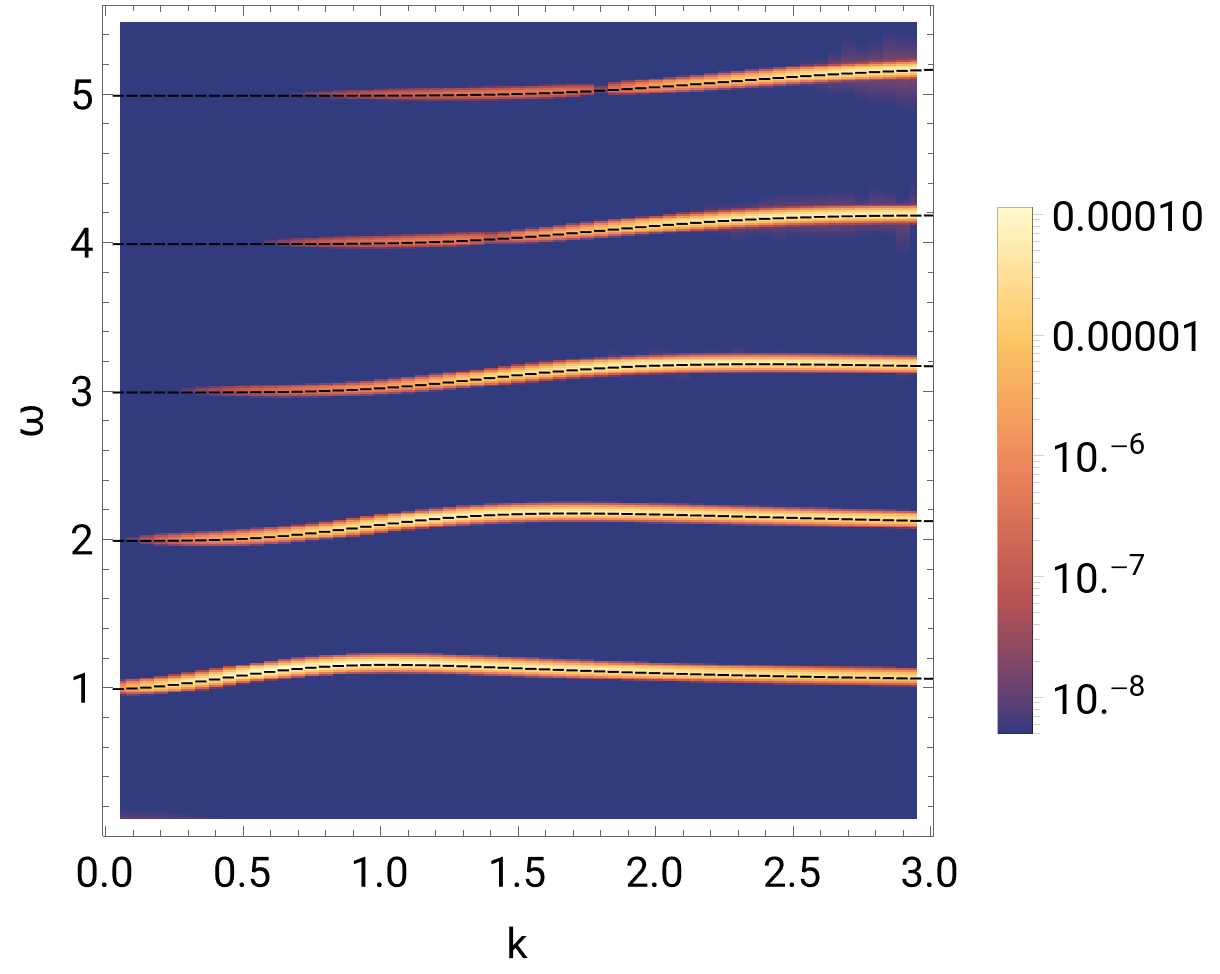}
        \caption{Absolute value of the numerically obtained temporal Fourier
        amplitudes of the electric potential as a function of frequency and
        perpendicular wavenumber. Dashed lines indicate the roots of the
        electrostatic ion-Bernstein dispersion relation.}
        \label{fig:adiabaticIBW}
        \vspace{0.5cm}
    \end{subfigure}
    \begin{subfigure}[t]{0.75\columnwidth}
        \centering
        \includegraphics[width=\linewidth]{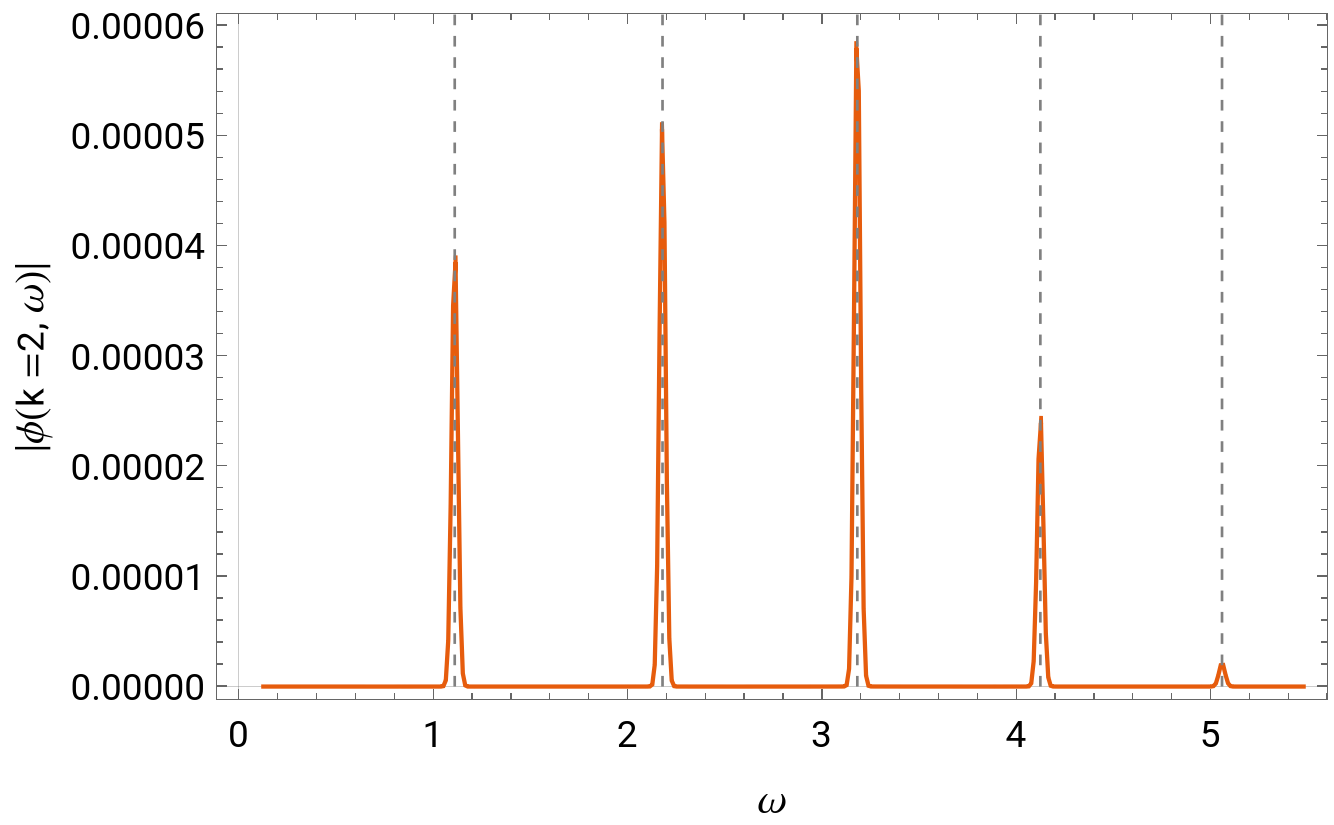}
        \caption{Spectral slice at fixed perpendicular wavenumber. The orange
        line shows the absolute value of the numerical Fourier amplitudes of
        the electric potential, while dashed vertical lines mark the
        corresponding dispersion-relation roots.}
        \label{fig:adiabaticIBWPeak}
    \end{subfigure}
    \caption{Standard dispersion-relation-based verification diagnostics for
    ion Bernstein waves. The comparison identifies the asymptotic mode
    frequencies but does not provide a pointwise-in-time reference solution.
    Simulations were performed on a $512 \times 33 \times 33$ grid, with one
    spatial and two velocity dimensions, on the domain
    $[0,20\pi]\times[-6,6]\times[-6,6]$ in normalized units. The time step was
    $\Delta t=0.02$, the final time was $t_{\mathrm{end}}=160\pi$, and the
    initial perturbation amplitude was $\lambda=10^{-6}$.}
    \label{fig:IBW}
\end{figure}

A common verification practice for such modes is to compare the numerical
spectrum with roots of the plasma dispersion relation in the undamped
$k_\parallel=0$ regime; see Fig.~\ref{fig:IBW}. Such comparisons are useful for
identifying asymptotic mode frequencies, but they do not provide a
pointwise-in-time reference solution. This limitation becomes particularly
important at finite $k_\parallel$, where parallel phase mixing damps the
Bernstein response. To address this issue, the present work generalizes our semi-analytical construction to plasmas with spatially homogeneous background quantities, namely a uniform background magnetic field and a Maxwellian equilibrium distribution.

The paper is organized as follows. First, we derive the distribution-function
response of a magnetized plasma to a prescribed single-mode electrostatic
potential. We then specialize to a plasma with kinetic ions and adiabatic
electrons, derive the electrostatic ion-density response for finite
$k_\parallel$, and compare the resulting semi-analytical solution with BSL6D
simulations. Subsequently, we discuss the singular $k_\parallel=0$ limit and
briefly outline how the response-function construction may be generalized to
fully electromagnetic scenarios.

\section{Magnetized electrostatic distribution-function response}

To derive the electrostatic distribution-function response of a uniformly
magnetized plasma, we consider a small single-mode perturbation and follow the
standard construction documented in the literature~\cite{stix, brambilla}. We
first decompose the distribution function as
\begin{equation}
    f(\bm{x},\bm{v},t)
    =
    f_0(\bm{v})+\lambda f_1(\bm{x},\bm{v},t)+O(\lambda^2),
    \qquad |\lambda|\ll1,
\end{equation}
with initial perturbation
\begin{equation}
    \label{perturbation}
    f_1(\bm{x},\bm{v},0)
    =
    f_0(\bm{v}) e^{i\bm{k}\cdot\bm{x}} .
\end{equation}
The stationary background is assumed to be Maxwellian,
\begin{equation}
    \label{maxwellian}
    f_0(\bm{v})
    =
    n_0 f_{\mathrm M}(\bm{v})
    =
    n_0 (2\pi v_{\mathrm{th}}^2)^{-3/2}
    \exp\left(-\frac{v^2}{2v_{\mathrm{th}}^2}\right),
\end{equation}
where $v_{\mathrm{th}} \coloneq \sqrt{T/m}$, and $m$, $n_0$, and $T$ denote the
particle mass, equilibrium density, and temperature, respectively. We further
assume a uniform background magnetic field
\begin{equation}
    \bm{B}=B\hat{\bm z}
\end{equation}
and neglect magnetic perturbations. To first order in $\lambda$, the Vlasov
equation becomes
\begin{equation}
    \label{linearVlasov}
    \partial_t f_1
    +\bm{v}\cdot\nabla f_1
    -\frac qm \nabla\phi\cdot\nabla_{\bm v}f_0
    +\frac qm \bm{v}\times\bm{B}\cdot\nabla_{\bm v}f_1
    =0 .
\end{equation}
In contrast to the unmagnetized case, the magnetic Lorentz term prevents a
purely algebraic solution for $f_1$ in Fourier space. We therefore separate the
perturbation into adiabatic and non-adiabatic contributions,
\begin{equation}
    \label{perturbationAnsatz}
    f_1(\bm{x},\bm{v},t)
    =
    h(\bm{x},\bm{v},t)
    -\frac qT \phi(\bm{x},t)f_0(\bm{v}) .
\end{equation}
Using $\nabla_{\bm v}f_0=-(m/T)\bm v f_0$, the terms proportional to
$\bm v\cdot\nabla\phi$ cancel, and Eq.~\eqref{linearVlasov} reduces to
\begin{equation}
    \label{nonAdPert}
    \partial_t h
    +\bm{v}\cdot\nabla h
    +\frac qm \bm{v}\times\bm{B}\cdot\nabla_{\bm v}h
    =
    \frac qT f_0(\bm{v})\partial_t\phi .
\end{equation}
The homogeneous part of Eq.~\eqref{nonAdPert} describes advection along
unperturbed particle orbits,
\begin{equation}
    \label{guidingCenter}
    \dot{\bm{x}}=\bm v,
    \quad
    \dot{\bm v}=\frac qm \bm v\times\bm B .
\end{equation}
Introducing the signed cyclotron frequency
\begin{equation}
    \Omega \coloneq \frac{qB}{m},
\end{equation}
and choosing the $x$-axis along $\bm k_\perp$, these orbits can be written as
\begin{equation}
    \bm{x}(t)
    =
    \bm{x}(0)+v_z\hat{\bm z}t
    +\frac{1}{\Omega}\hat{\bm z}\times
    \left[\bm v_\perp(t)-\bm v_\perp(0)\right],
\end{equation}
where
\begin{equation}
    \bm v_\perp(t)
    =
    v_\perp
    \begin{pmatrix}
        \cos(\Omega t+\alpha)\\
        -\sin(\Omega t+\alpha)\\
        0
    \end{pmatrix}.
\end{equation}
Here, $\alpha$ is defined by $\bm{k}_\perp \cdot \bm{v}_\perp(0) = k_\perp v_\perp \cos\alpha$. The single mode propagator associated with Eq.~\eqref{nonAdPert}, which will be used below to solve for $h$, is then given by
\begin{equation}
    \label{greensFunctionA}
    G^+(\bm{x},\bm v,t)
    =
    \theta(t)\exp\left[i\bm k\cdot\bm x(-t)\right],
\end{equation}
where $\theta$ denotes the Heaviside function. Since
\begin{equation}
    \bm{k}\cdot\bm{x}(-t)
    =
    \bm{k}\cdot\bm{x}
    -k_zv_z t
    -\frac{k_\perp v_\perp}{\Omega}\sin(\Omega t+\alpha)
    +\frac{k_\perp v_\perp}{\Omega}\sin\alpha ,
\end{equation}
the Jacobi-Anger expansions
\begin{align}
    e^{-i \frac{v_\perp k_\perp}{\Omega} \sin(\Omega t+\alpha)}
    &=
    \sum_{p=-\infty}^{\infty}
    J_p\left(\frac{v_\perp k_\perp}{\Omega}\right)
    e^{-ip(\Omega t+\alpha)}, \\
    e^{i \frac{v_\perp k_\perp}{\Omega} \sin\alpha}
    &=
    \sum_{m=-\infty}^{\infty}
    J_m\left(\frac{v_\perp k_\perp}{\Omega}\right)
    e^{im\alpha}
\end{align}
yield, for $t>0$,
\begin{align}
    \label{greensFunctionB}
    &G^+(\bm{x},\bm v,t) = \notag \\
    &\quad e^{i\bm{k}\cdot\bm{x}}
    \sum_{m,p=-\infty}^{\infty}
    J_m\left(\frac{v_\perp k_\perp}{\Omega}\right)
    J_p\left(\frac{v_\perp k_\perp}{\Omega}\right)
    e^{i(m-p)\alpha}
    e^{-i(\Omega p+k_zv_z)t}.
\end{align}
Because the problem is linear and the background is homogeneous, the initial
condition~\eqref{perturbation} evolves independently in the Fourier mode
$\bm k$. With a slight abuse of notation, we therefore write
\begin{align}
    f_1(\bm{x},\bm v,t)
    &=
    f_1(\bm v,t)e^{i\bm k\cdot\bm x}, \\
    h(\bm{x},\bm v,t)
    &=
    h(\bm v,t)e^{i\bm k\cdot\bm x}, \\
    \phi(\bm{x},t)
    &=
    \phi(t)e^{i\bm k\cdot\bm x}.
\end{align}
For the causal response, we define
\begin{equation}
    f_1^+=\theta(t)f_1,
    \quad
    h^+=\theta(t)h,
    \quad
    \phi^+=\theta(t)\phi, 
\end{equation}
and use the temporal Fourier convention
\begin{equation}
    \label{FT}
    \hat g(\omega)
    \coloneq
    \int_{-\infty}^{\infty}
    dt\, e^{i\omega t}g(t),
    \quad
    g(t)
    \coloneq
    \int_{-\infty}^{\infty}
    \frac{d\omega}{2\pi}\, e^{-i\omega t}\hat g(\omega).
\end{equation}
Assuming $\operatorname{Im}\omega>0$ to ensure convergence of the Fourier
integrals, Eq.~\eqref{nonAdPert} gives
\begin{align}
    \label{linearVlasovFou}
    &-i(\omega-\bm k\cdot\bm v)\hat h^+
    +\frac qm \bm v\times\bm B\cdot\nabla_{\bm v}\hat h^+
    \notag \\
    &\qquad\qquad\qquad\qquad
    =
    f_1(\bm v,t=0)
    -i\omega\frac qT f_0(\bm v)\hat\phi^+ .
\end{align}
The term $f_1(\bm v,t=0)$ arises from the discontinuities of $h^+$ and $\phi^+$
at $t=0$, with the corresponding boundary contributions combining according to
Eq.~\eqref{perturbationAnsatz}. The solution of Eq.~\eqref{linearVlasovFou} can
be expressed in terms of the causal propagator as
\begin{equation}
    \label{greensSol}
    \hat h^+
    =
    \hat G^+(\omega)
    \left[
        f_1(\bm v,0)
        -i\omega\frac qT f_0(\bm v)\hat\phi^+
    \right],
\end{equation}
where Eq.~\eqref{greensFunctionB} yields, for $\operatorname{Im}\omega>0$,
\begin{equation}
    \label{greensFunctionC}
    \hat G^+(\omega)
    =
    \sum_{m,p=-\infty}^{\infty}
    J_m\left(\frac{v_\perp k_\perp}{\Omega}\right)
    J_p\left(\frac{v_\perp k_\perp}{\Omega}\right)
    \frac{i e^{i(m-p)\alpha}}
    {\omega-k_zv_z-\Omega p}.
\end{equation}
Consequently,
\begin{equation}
    \label{analyticfrespCompact}
    \hat f_1^+
    =
    \hat G^+(\omega)
    \left[
        f_1(\bm v,0)
        -i\omega\frac qT f_0(\bm v)\hat\phi^+
    \right]
    -
    \frac qT f_0(\bm v)\hat\phi^+ .
\end{equation}
For the normalized initial perturbation~\eqref{perturbation}, this becomes
\begin{align}
    \label{analyticfresp}
    \hat{f}_1^+
    =
    &\sum_{m,p=-\infty}^{\infty}
    J_m\left(\frac{v_\perp k_\perp}{\Omega}\right)
    J_p\left(\frac{v_\perp k_\perp}{\Omega}\right)
    \frac{i e^{i(m-p)\alpha}}
    {\omega-k_z v_z-\Omega p}
    \notag \\
    &\times
    \left(
        1-\frac{i\omega q}{T}\hat{\phi}^+
    \right)
    f_0(\bm v)
    -
    \frac{q}{T}\hat{\phi}^+ f_0(\bm v).
\end{align}

\section{Magnetized electrostatic density response}

\subsection{Density response for a given electrostatic potential}

The plasma density response is obtained by integrating Eq. \eqref{analyticfresp} over velocity space,
\begin{equation}
    \hat n_1^+(\bm{k},\omega)
    \coloneq
    \int d^3v\, \hat f_1^+
    =
    \int_0^{2\pi} d\alpha
    \int_0^\infty dv_\perp\, v_\perp
    \int_{-\infty}^{\infty} dv_z\, \hat f_1^+ ,
\end{equation}
where
\begin{equation}
    \hat f_1^+
    =
    \hat h^+
    -
    \frac qT \hat\phi^+ f_0(\bm v).
\end{equation}
We first evaluate the non-adiabatic contribution $\hat h^+$. Using
\begin{equation}
    \label{alphaIntegral}
    \int_0^{2\pi} d\alpha\, e^{i(m-p)\alpha}
    =
    2\pi\delta_{m,p},
    \qquad m,p\in\mathbb Z,
\end{equation}
the gyro-phase integration gives
\begin{align}
    \label{alphaIntegration}
    \int_0^{2\pi} d\alpha\, \hat h^+
    =
    \sum_{p=-\infty}^{\infty}
    J_p^2\left(\frac{v_\perp k_\perp}{\Omega}\right)
    \frac{2\pi i n_0}
    {\omega-k_zv_z-\Omega p}
    \notag \\
    \times
    \left(
        1-\frac{i\omega q}{T}\hat\phi^+
    \right)
    f_{\mathrm M}(v_\perp)f_{\mathrm M}(v_z),
\end{align}
where
\begin{equation}
    f_{\mathrm M}(v_\perp)
    \coloneq
    \frac{1}{2\pi v_{\mathrm{th}}^2}
    e^{-\frac{v_\perp^2}{2v_{\mathrm{th}}^2}},
    \qquad
    f_{\mathrm M}(v_z)
    \coloneq
    \frac{1}{\sqrt{2\pi}v_{\mathrm{th}}}
    e^{-\frac{v_z^2}{2v_{\mathrm{th}}^2}}.
\end{equation}
The $v_\perp$ integration can be evaluated using the identity
\begin{equation}
    \label{vperpIntegral}
    \int_0^\infty dv_\perp\, v_\perp
    J_p^2\left(\frac{v_\perp k_\perp}{\Omega}\right)
    f_{\mathrm M}(v_\perp)
    =
    \frac{1}{2\pi}
    \exp\left(-\frac{k_\perp^2v_{\mathrm{th}}^2}{\Omega^2}\right)
    I_p\left(\frac{k_\perp^2v_{\mathrm{th}}^2}{\Omega^2}\right),
\end{equation}
where $I_p$ denotes the modified Bessel function of the first kind \cite{gradshteyn}. We define the positive thermal Larmor radius
\begin{equation}
    \rho
    \coloneq
    \frac{v_{\mathrm{th}}}{|\Omega|}
\end{equation}
and
\begin{equation}
    \Gamma_p(\rho^2k_\perp^2)
    \coloneq
    e^{-\rho^2k_\perp^2}I_p(\rho^2k_\perp^2).
\end{equation}
Since $I_{-p}=I_p$, the harmonic sum can be written as
\begin{align}
    &\int_0^{2\pi} d\alpha
    \int_0^\infty dv_\perp\, v_\perp\, \hat h^+
    =
    i n_0
    \left(
        1-\frac{i\omega q}{T}\hat\phi^+
    \right)
    \notag \\
    &\times
    \left[
    \frac{\Gamma_0 f_{\mathrm M}(v_z)}
    {\omega-k_zv_z}
    +
    \sum_{p=1}^{\infty}
    \left(
    \frac{\Gamma_p f_{\mathrm M}(v_z)}
    {\omega-k_zv_z-\Omega p}
    +
    \frac{\Gamma_p f_{\mathrm M}(v_z)}
    {\omega-k_zv_z+\Omega p}
    \right)
    \right].
\end{align}
For $k_z >0$, the remaining $v_z$ integration can be expressed in terms of the plasma dispersion function \cite{nrl},
\begin{equation}
    \label{zFunctionA}
    Z(\zeta)
    \coloneq
    \frac{1}{\sqrt{\pi}}
    \int_{-\infty}^{\infty} dx\,
    \frac{e^{-x^2}}{x-\zeta},
    \qquad
    \operatorname{Im}\zeta>0,
\end{equation}
so that
\begin{equation}
    \label{zFunctionResult}
    \int_{-\infty}^{\infty} dv_z\,
    \frac{f_{\mathrm M}(v_z)}
    {\omega-k_zv_z-\Omega p}
    =
    -
    \frac{1}{k_z\sqrt{2}v_{\mathrm{th}}}
    Z\left(
        \frac{\omega-\Omega p}
        {k_z\sqrt{2}v_{\mathrm{th}}}
    \right).
\end{equation}
Rather than evaluating the integral definition \eqref{zFunctionA} directly, we use the entire analytic continuation
\begin{equation}
    \label{zfunctionB}
    Z(\zeta)
    =
    i\sqrt{\pi}e^{-\zeta^2}
    -2D(\zeta),
    \qquad
    D(\zeta)
    \coloneq
    e^{-\zeta^2}\int_0^\zeta ds\, e^{s^2},
\end{equation}
where $D(\zeta)$ is the Dawson function. In summary,
\begin{align}
    &\int d^3v\, \hat h^+
    =
    -\frac{i n_0}{k_z\sqrt{2}v_{\mathrm{th}}}
    \left(
        1-\frac{i\omega q}{T}\hat\phi^+
    \right)
    \notag \\
    &\quad \times
    \left[
    Z\left(
        \frac{\omega}
        {k_z\sqrt{2}v_{\mathrm{th}}}
    \right)\Gamma_0
    +
    \sum_{p=1}^{\infty}
    \left(
    Z\left(
        \frac{\omega-\Omega p}
        {k_z\sqrt{2}v_{\mathrm{th}}}
    \right)
    +
    Z\left(
        \frac{\omega+\Omega p}
        {k_z\sqrt{2}v_{\mathrm{th}}}
    \right)
    \right)\Gamma_p
    \right].
\end{align}
Defining 
\begin{align}
    \label{greensFctDensity}
    \hat H_l^+(\omega, \bm{k})
    \coloneq &
    -
    \frac{1}{k_z\sqrt{2}v_{\mathrm{th}}}
    \Bigg[
    Z\left(
        \frac{\omega}
        {k_z\sqrt{2}v_{\mathrm{th}}}
    \right)\Gamma_0
    \notag \\
    \qquad
    &+
    \sum_{p=1}^{l}
    \left(
    Z\left(
        \frac{\omega-\Omega p}
        {k_z\sqrt{2}v_{\mathrm{th}}}
    \right)
    +
    Z\left(
        \frac{\omega+\Omega p}
        {k_z\sqrt{2}v_{\mathrm{th}}}
    \right)
    \right)\Gamma_p
    \Bigg]
\end{align}
and including the adiabatic part of $f_1^+$, one arrives at
\begin{equation}
    \label{nRespEq}
    \hat{n}_1^+(\omega, \bm{k})
    =
    i n_0
    \left(
        1 - \frac{i \omega q}{T} \hat{\phi}^+
    \right)
    \hat H_\infty^+(\omega, \bm{k})
    -
    \frac{q n_0}{T}\hat{\phi}^+ .
\end{equation}

\subsection{Density response with adiabatic electrons}

We now close the system by imposing quasineutrality and an adiabatic
electron response. Under these assumptions, the electrostatic potential is
given by~\cite{Knorr:1970}
\begin{equation}
    \label{electricPotential}
    \hat\phi^+(\omega,\bm{k})
    =
    \frac{T_e}{q n_0}
    \hat n_1^+(\omega,\bm{k}),
\end{equation}
where $T_e$ is the electron temperature. Introducing the temperature ratio
\begin{equation}
    \tau \coloneq \frac{T_e}{T},
\end{equation}
Eq.~\eqref{nRespEq} becomes
\begin{equation}
    \hat n_1^+
    =
    i n_0 \hat H_\infty^+
    +
    \tau \omega \hat H_\infty^+ \hat n_1^+
    -
    \tau \hat n_1^+ ,
\end{equation}
and solving for the density response yields
\begin{equation}
    \label{nResp}
    \hat n_1^+(\omega,\bm{k})
    =
    \frac{
        i n_0 \hat H_\infty^+(\omega,\bm{k})
    }{
        1+\tau-\tau\omega\hat H_\infty^+(\omega,\bm{k})
    } .
\end{equation}
For finite $k_z$, the cyclotron-harmonic singularities are regularized by the
parallel velocity integration, and the response function is regular on the real
$\omega$-axis. In the strictly perpendicular limit $k_z=0$, however, this is no
longer the case, since
\begin{equation}
    \label{kzLimitZ}
    \lim_{k_z\to0}
    \frac{1}{k_z\sqrt{2}v_{\mathrm{th}}}
    Z\left(
        \frac{\omega-\Omega p}
        {k_z\sqrt{2}v_{\mathrm{th}}}
    \right)
    =
    -\frac{1}{\omega-\Omega p},
\end{equation}
and hence
\begin{equation}
    \label{HkzZero}
    \hat H_\infty^+(\omega,k_z = 0)
    =
    \sum_{p=-\infty}^{\infty}
    \frac{\Gamma_p}{\omega-\Omega p}.
\end{equation}
Thus, for $k_z=0$, the poles of the response are determined by the zeros of
the denominator in Eq.~\eqref{nResp}, i.e. by
\begin{equation}
    \label{bernsteinDispersion}
    1+\tau
    -
    \tau\omega
    \sum_{p=-\infty}^{\infty}
    \frac{\Gamma_p}{\omega-\Omega p}
    =
    0 .
\end{equation}
These roots lie on the real axis, so that a direct numerical inverse Fourier
transform along the real-frequency axis is no longer straightforward. In the
following sections, we first focus on the damped finite-$k_z$ response and
return to the strictly perpendicular limit afterwards.

\section{Time-domain benchmark for the finite-$k_z$ density response}

There are three sources of numerical error in the Fourier inversion of the
density response~\eqref{nResp}: truncation of the frequency interval,
quadrature error, and truncation of the cyclotron-harmonic sum. We first recall
the response-symmetrization procedure introduced in our previous
work~\cite{isw}, which controls both the frequency-truncation and
quadrature errors. We then discuss the controlled truncation of the harmonic
sum and derive an estimate for the total error.

\subsubsection{Response symmetrization}

\begin{figure}[ht]
    \centering
    \begin{subfigure}[t]{0.75\columnwidth}
        \centering
        \includegraphics[width=\textwidth]{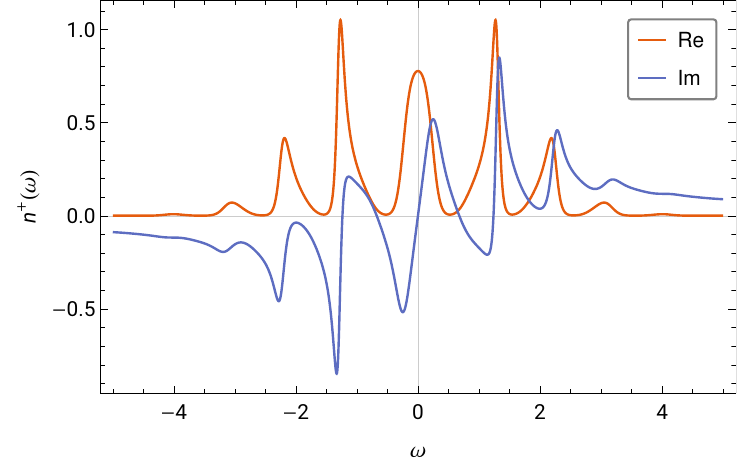}
        \caption{Spectrum of the causal density response for $k_\perp=1$,
        $k_z=0.15$, and harmonic truncation at $p=5$. The slowly decaying
        contribution generated by the discontinuity at $t=0$ leads to
        Gibbs-type oscillations after numerical Fourier inversion.}
        \label{fig:nRespSpectrumP5}
        \vspace{0.25cm}
    \end{subfigure}
    \begin{subfigure}[t]{0.75\columnwidth}
        \centering
        \includegraphics[width=\textwidth]{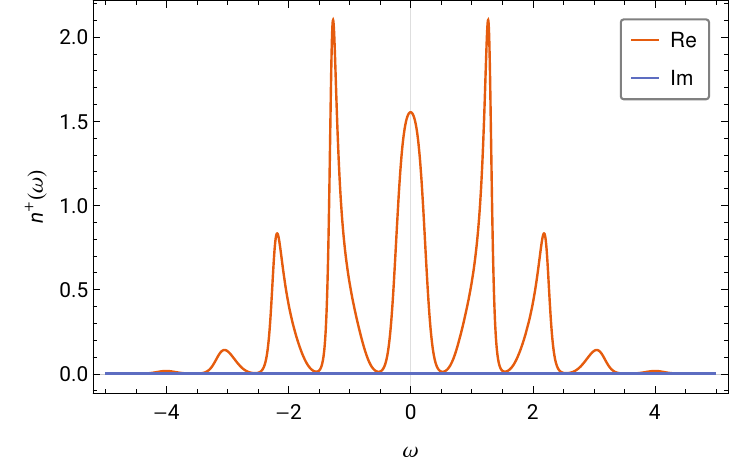}
        \caption{Spectrum of the symmetrized density response, again for
        $k_\perp=1$, $k_z=0.15$, and harmonic truncation at $p=5$. Removing the
        discontinuity at $t=0$ eliminates the spectral tail and yields a rapidly
        decaying spectrum.}
        \label{fig:nRespSpectrumSymP5}
    \end{subfigure}
    \caption{Causal and symmetrized density-response spectra, with the
    cyclotron-harmonic sum truncated at $p=5$.}
    \label{fig:nRespSpectrum}
\end{figure}

\begin{figure}[ht]
    \centering
    \begin{subfigure}[t]{0.75\columnwidth}
        \centering
        \includegraphics[width=\textwidth]{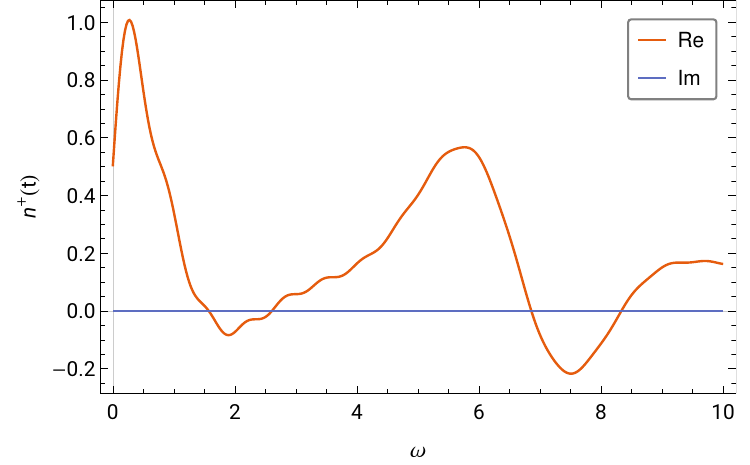}
        \caption{Causal density response. The slowly decaying spectral tail,
        originating from the discontinuity at $t=0$, introduces Gibbs-type
        oscillations in the reconstructed solution.}
        \label{fig:nRespAsym}
        \vspace{0.25cm}
    \end{subfigure}
    \begin{subfigure}[ht]{0.75\columnwidth}
        \centering
        \includegraphics[width=\textwidth]{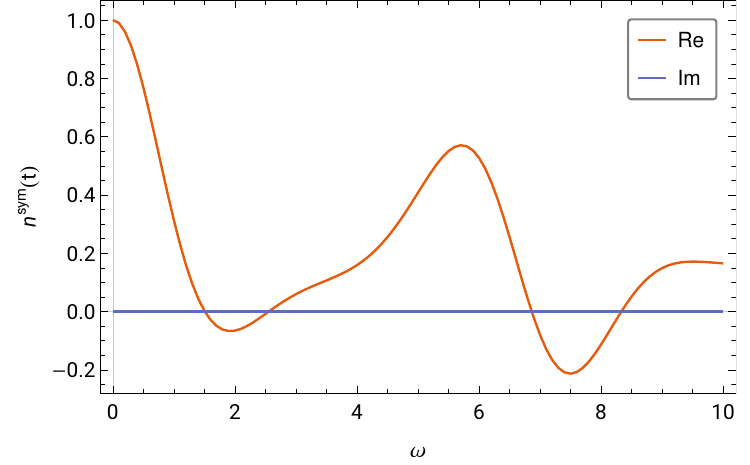}
        \caption{Symmetrized density response. Eliminating the spectral tail
        removes the Gibbs-type oscillations.}
        \label{fig:nRespSym}
    \end{subfigure}
    \caption{Effect of time symmetrization on the finite-$k_z$ density
    response. The causal response defines the initial-value problem, while the
    symmetrized response provides the numerically useful representation for
    reconstructing the time-domain benchmark.}
    \label{fig:nResp}
\end{figure}

Due to its discontinuity at $t=0$, the spectrum of the causal density response,
\begin{equation}
    n_1^+(\bm{k},t)
    \coloneq
    \theta(t)n_1(\bm{k},t),
\end{equation}
contains a slowly decaying spectral tail; see Fig.~\ref{fig:nRespSpectrum}.
This is a generic feature of causal responses, as can be seen from the Fourier
transform of the Heaviside step function,
\begin{equation}
   \int_{-\infty}^{\infty} dt\, e^{i\omega t}\theta(t)
   =
   \pi\delta(\omega)
   +
   i\,\operatorname{p.v.}\left(\frac{1}{\omega}\right),
\end{equation}
where $\operatorname{p.v.}$ denotes the Cauchy principal value. Because of this
slow spectral decay, truncating the Fourier integration interval inevitably
introduces Gibbs-type oscillations in the reconstructed time signal
(Fig.~\ref{fig:nResp}), rendering the resulting solution unsuitable for
high-precision code verification.

The spectral tail, however, is not a physical property of the linearized Vlasov
dynamics. Rather, it is caused by replacing the full time history by a one-sided
signal that contains the initial condition at $t=0$ and is set to zero for
$t<0$. The idea of the symmetrization procedure is to reconstruct the missing
$t<0$ branch by evolving the initial condition backward in time and adding it
to the causal $t>0$ solution, thereby removing the temporal discontinuity at
$t=0$. For initial conditions that are compatible with a smooth, compactly
supported time evolution of the distribution function, the symmetrized
frequency spectrum decays faster than any inverse polynomial~\cite{Stein:2011}.
This rapid decay makes an accurate truncation of the numerical Fourier interval
possible; see Fig.~\ref{fig:nRespSpectrum}.

In previous work~\cite{isw}, we showed that, for initial conditions with
definite time parity, this construction reduces to a symmetrization or
antisymmetrization of the frequency spectrum obtained from the causal $t>0$
solution. For the pure density perturbation in Eq.~\eqref{perturbation}, which
has even time parity, the spectrum must be symmetrized. Thus, we define
\begin{equation}
    \label{symmetrizationDefinition}
    \hat n_{1}^{\mathrm{sym}}(\omega,\bm{k})
    \coloneq
    \lim_{\epsilon\to0^+}
    \left[
        \hat n_1^+(\omega+i\epsilon,\bm{k})
        +
        \hat n_1^+(-\omega+i\epsilon,\bm{k})
    \right],
    \qquad
    \omega\in\mathbb R .
\end{equation}
Setting $T_e=T$ and using Eq.~\eqref{nResp}, the finite-$k_z$ symmetrized
density response becomes
\begin{equation}
    \label{nSym}
    \hat n_1^{\mathrm{sym}}(\omega,\bm{k})
    =
    \frac{
        2i n_0\,\hat H_\infty^{\mathrm{sym}}(\omega,\bm{k})
    }{
        \left[
            2-\omega\hat H^+_\infty(\omega,\bm{k})
        \right]
        \left[
            2+\omega\hat H^+_\infty(-\omega,\bm{k})
        \right]
    },
\end{equation}
where
\begin{equation}
    \hat H_l^{\mathrm{sym}}(\omega,\bm{k})
    \coloneq
    \hat H_l^+(\omega,\bm{k})
    +
    \hat H_l^+(-\omega,\bm{k}) .
\end{equation}

\subsubsection{Truncation of the cyclotron-harmonic sum}

With the symmetrized density spectrum at hand, we can derive a bound for the
pointwise error of the Fourier-inverted time-domain density response. To this
end, we introduce the truncated symmetrized response
\begin{equation}
    \label{nSymTruncated}
    \hat n_{1,l}^{\mathrm{sym}}(\omega,\bm{k})
    \coloneq
    \frac{
        2i n_0
        \hat H_l^{\mathrm{sym}}(\omega,\bm{k})
    }{
        \hat Q_l(\omega,\bm{k})
    },
\end{equation}
where
\begin{equation}
    \label{Qdef}
    \hat Q_l(\omega,\bm{k})
    \coloneq
    \left[
        2-\omega\hat H_l^+(\omega,\bm{k})
    \right]
    \left[
        2+\omega\hat H_l^+(-\omega,\bm{k})
    \right].
\end{equation}
For the moment, we neglect the quadrature error and treat the Fourier integral
as continuous. Denoting the numerical cutoff frequency by $a$, the symmetrized
time-domain response is approximated by
\begin{equation}
    \label{symmetrizedInverseFourier}
    n_{1,l,a}^{\mathrm{sym}}(\bm{k},t)
    \coloneq
    \int_0^a
    \frac{d\omega}{\pi}\,
    \cos(\omega t)\,
    \hat n_{1,l}^{\mathrm{sym}}(\omega,\bm{k}) .
\end{equation}
The pointwise inversion error satisfies
\begin{align}
    \label{numericalError}
    \epsilon_{l,a}(t)
    &\coloneq
    \left|
    n_{1,\infty,\infty}^{\mathrm{sym}}(\bm{k},t)
    -
    n_{1,l,a}^{\mathrm{sym}}(\bm{k},t)
    \right|
    \nonumber\\
    &\leq
    \frac{1}{\pi}
    \int_a^\infty
    d\omega\,
    \left|
        \hat n_{1,\infty}^{\mathrm{sym}}(\omega,\bm{k})
    \right|
    \nonumber\\
    &\quad
    +
    \frac{1}{\pi}
    \int_0^a
    d\omega\,
    \left|
        \hat n_{1,\infty}^{\mathrm{sym}}(\omega,\bm{k})
        -
        \hat n_{1,l}^{\mathrm{sym}}(\omega,\bm{k})
    \right| ,
\end{align}
where the first term is the frequency-cutoff error, while the second term is the
error caused by truncating the cyclotron-harmonic sum inside the retained
frequency interval. For finite $k_z$, the denominator $\hat Q_\infty$ has no
real zeros and is bounded away from zero. We therefore define
\begin{equation}
    \label{cQinfDef}
    c_1
    \coloneq
    \sup_{\omega\in\mathbb R}
    \left|
        2 n_0 \hat Q_\infty(\omega,\bm{k})^{-1}
    \right|.
\end{equation}

\begin{figure}[h]
    \centering
    \begin{subfigure}[t]{0.75\columnwidth}
        \centering
        \includegraphics[width=\textwidth]{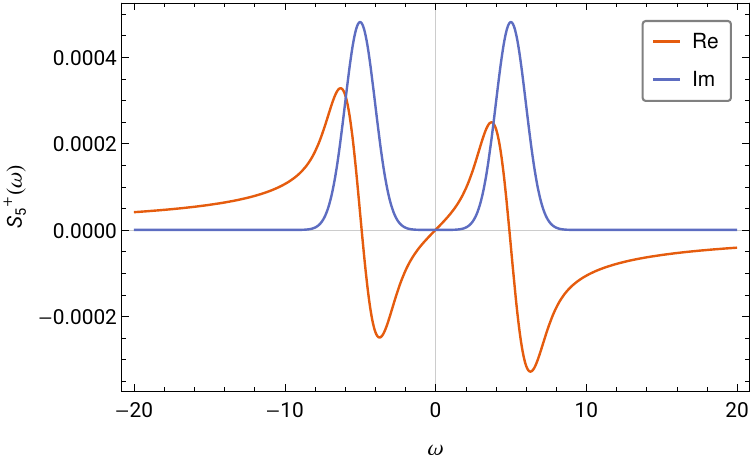}
        \caption{Frequency dependence of the $p=5$ cyclotron-harmonic
        contribution, denoted by $S_5^+$, to $\hat H^+$ for
        $k_\perp=k_z=1$.}
        \label{fig:harmonicContributionP5}
        \vspace{0.5cm}
    \end{subfigure}
    \begin{subfigure}[t]{0.75\columnwidth}
        \centering
        \includegraphics[width=\textwidth]{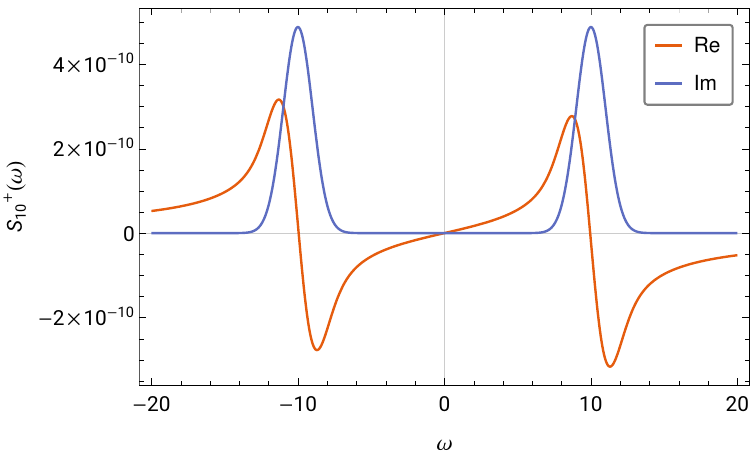}
        \caption{Frequency dependence of the $p=10$ cyclotron-harmonic
        contribution, denoted by $S_{10}^+$, to $\hat H^+$ for
        $k_\perp=k_z=1$.}
        \label{fig:harmonicContributionP10}
    \end{subfigure}
    \caption{Individual cyclotron-harmonic contributions to the magnetized
    density response function. Higher harmonics are centered around increasingly
    large cyclotron frequencies $\omega\simeq \pm p\Omega$ and are increasingly
    suppressed by the Bessel weights $\Gamma_p(\rho^2k_\perp^2)$.}
    \label{fig:harmonicContributions}
\end{figure}

By inspection, it follows that $c_1 \leq 2 n_0$. Using Eq.~\eqref{nSymTruncated}, the frequency-cutoff contribution can be bounded by
\begin{align}
    \int_a^\infty
    d\omega\,
    \left|
        \hat n_{1,\infty}^{\mathrm{sym}}
    \right|
    &\leq
    c_1
    \int_a^\infty
    d\omega\,
    \left|
        \hat H_\infty^{\mathrm{sym}}
    \right|
    \nonumber\\
    &\leq
    c_1
    \left[
    \int_a^\infty
    d\omega\,
    \left|
        \hat H_l^{\mathrm{sym}}
    \right|
    +
    \int_a^\infty
    d\omega\,
    \left|
        \Delta\hat H_l^{\mathrm{sym}}
    \right|
    \right],
\end{align}
where $ \Delta\hat H_l^{\mathrm{sym}}
    \coloneq
    \hat H_\infty^{\mathrm{sym}}
    -
    \hat H_l^{\mathrm{sym}} $. From Eq.~\eqref{greensFctDensity}, one finds that, for $p\geq1$, the
symmetrized $p$-th harmonic contribution to $\hat H_l^{\mathrm{sym}}$ is
\begin{equation}
    -\frac{i\sqrt{2\pi}}{|k_z|v_{\mathrm{th}}}
    \Gamma_p
    \left[
        e^{-\frac{(\omega-p\Omega)^2}{2k_z^2v_{\mathrm{th}}^2}}
        +
        e^{-\frac{(\omega+p\Omega)^2}{2k_z^2v_{\mathrm{th}}^2}}
    \right].
\end{equation}
Thus, the symmetrized harmonic contributions are Gaussian packets centered at
the cyclotron frequencies $\omega=\pm p\Omega$, with amplitudes controlled by
the Bessel weights $\Gamma_p$; see Fig.~\ref{fig:harmonicContributions}. The
$p=0$ contribution is the corresponding single Gaussian centered at
$\omega=0$. Hence,
\begin{align}
    \label{retainedTailBound}
    R_l(a)
    &\coloneq
    \int_a^\infty
    d\omega\,
    \left|
        \hat H_l^{\mathrm{sym}}(\omega,\bm{k})
    \right|
    \nonumber\\
    &\leq
    \pi\Gamma_0
    \operatorname{erfc}
    \left(
        \frac{a}
        {\sqrt{2}\,|k_z|v_{\mathrm{th}}}
    \right) + r ,
\end{align}
where 
\begin{equation}
    r \coloneq \pi
    \sum_{p=1}^{l}
    \Gamma_p
    \left[
    \operatorname{erfc}
    \left(
        \frac{a-p|\Omega|}
        {\sqrt{2}\,|k_z|v_{\mathrm{th}}}
    \right)
    +
    \operatorname{erfc}
    \left(
        \frac{a+p|\Omega|}
        {\sqrt{2}\,|k_z|v_{\mathrm{th}}}
    \right) \right ].
\end{equation}
When $a>l|\Omega|$ and the Gaussian packets are sufficiently separated, this
bound is dominated by the packet centered at the largest retained cyclotron
frequency. It remains to control the contribution from the neglected harmonics,
\begin{align}
    \label{DeltaHsym}
    \Delta\hat H_l^{\mathrm{sym}}(\omega,\bm{k})
    =
    -\frac{i\sqrt{2\pi}}{|k_z|v_{\mathrm{th}}}
    \sum_{p=l+1}^{\infty}
    \Gamma_p
    \left[
        e^{-\frac{(\omega-p\Omega)^2}{2k_z^2v_{\mathrm{th}}^2}}
        +
        e^{-\frac{(\omega+p\Omega)^2}{2k_z^2v_{\mathrm{th}}^2}}
    \right].
\end{align}
Let
\begin{equation}
    x\coloneq \rho^2k_\perp^2 .
\end{equation}
Integrating the Gaussian bound in Eq.~\eqref{DeltaHsym} over the real line gives
the simple estimate
\begin{equation}
    \label{HtailBoundA}
    \int_a^\infty d\omega\,
    \left|
        \Delta\hat H_l^{\mathrm{sym}}
    \right|
    \leq
    4 \pi
    \sum_{p=l+1}^{\infty}
    \Gamma_p(x).
\end{equation}
To bound the infinite series over the Bessel weights, we use the recurrence
relation~\cite{gradshteyn}
\begin{equation}
    I_p(x)-I_{p+2}(x)
    =
    \frac{2(p+1)}{x}I_{p+1}(x).
\end{equation}
Since $I_p(x)>0$ for $x>0$, we find
\begin{equation}
    \label{besselRatioBound}
    \frac{\Gamma_{p+1}(x)}{\Gamma_p(x)}
    =
    \frac{I_{p+1}(x)}{I_p(x)}
    =
    \frac{x}{2(p+1)}
    \left(
        1-\frac{I_{p+2}(x)}{I_p(x)}
    \right)
    \leq
    \frac{x}{2(p+1)} .
\end{equation}
Hence, for $p\geq l+1$,
\begin{equation}
    \frac{\Gamma_{p+1}(x)}{\Gamma_p(x)}
    \leq
    \frac{x}{2(l+2)}
    \eqqcolon q_l .
\end{equation}
If $q_l<1$, which can always be achieved by choosing $l$ sufficiently large,
the harmonic tail is bounded by the geometric series
\begin{equation}
    \label{gammaTailBound}
    \sum_{p=l+1}^{\infty}
    \Gamma_p(x)
    \leq
    \frac{\Gamma_{l+1}(x)}{1-q_l}
    =
    \frac{\Gamma_{l+1}(x)}
    {1-x/[2(l+2)]}.
\end{equation}
We denote this bound by
\begin{equation}
    \label{etaDef}
    \eta_l
    \coloneq
    \frac{
        \Gamma_{l+1}(\rho^2k_\perp^2)
    }{
        1-\rho^2k_\perp^2/[2(l+2)]
    } ,
\end{equation}
so that
\begin{equation}
    \label{DeltaHTailBoundFinal}
    \int_a^\infty d\omega\,
    \left|
        \Delta\hat H_l^{\mathrm{sym}}
    \right|
    \leq
    4\pi\eta_l .
\end{equation}

To make the decay of $\eta_l$ explicit, we bound $\Gamma_{l+1}(x)$. Using the standard series representation
\begin{equation}
    \label{BesselI}
    I_n(x)
    =
    \sum_{m=0}^{\infty} T_m^n,
    \qquad
    T_m^n
    \coloneq
    \frac{1}{m!(m+n)!}
    \left(\frac{x}{2}\right)^{2m+n},
\end{equation}
the ratio of successive terms satisfies
\begin{equation}
    \frac{T_{m+1}^n}{T_m^n}
    =
    \frac{x^2}{4(m+1)(m+n+1)}
    \leq
    \frac{x^2}{4(n+1)}
    \eqqcolon r_n .
\end{equation}
Thus, for $r_n<1$,
\begin{equation}
    I_n(x)
    \leq
    \frac{1}{1-r_n}
    \frac{1}{n!}
    \left(
        \frac{x}{2}
    \right)^n .
\end{equation}
Using Stirling's lower bound,
\begin{equation}
    n!
    \geq
    \sqrt{2\pi n}
    \left(
        \frac{n}{e}
    \right)^n ,
\end{equation}
we obtain
\begin{equation}
    \label{GammaStirlingBound}
    \Gamma_n(x)
    \leq
    \frac{e^{-x}}{1-r_n}
    \frac{1}{\sqrt{2\pi n}}
    \left(
        \frac{ex}{2n}
    \right)^n .
\end{equation}
For $n=l+1$, this gives
\begin{equation}
    \label{GammaLp1Stirling}
    \Gamma_{l+1}(x)
    \leq
    \frac{e^{-x}}{1-r_{l+1}}
    \frac{1}{\sqrt{2\pi(l+1)}}
    \left(
        \frac{ex}{2(l+1)}
    \right)^{l+1},
\end{equation}
provided $r_{l+1}=x^2/[4(l+2)]<1$. Hence, for fixed $k_\perp$, the harmonic
tail decays super-exponentially in $l$. We finally estimate the second contribution in Eq.~\eqref{numericalError},
\begin{equation}
    \hat n_{1,\infty}^{\mathrm{sym}}
    -
    \hat n_{1,l}^{\mathrm{sym}}
    =
    2i n_0
    \left[
        \frac{\hat H_\infty^{\mathrm{sym}}}{\hat Q_\infty}
        -
        \frac{\hat H_l^{\mathrm{sym}}}{\hat Q_l}
    \right].
\end{equation}
Equivalently,
\begin{equation}
    \label{secondErrorContribution}
    \hat n_{1,\infty}^{\mathrm{sym}}
    -
    \hat n_{1,l}^{\mathrm{sym}}
    =
    2i n_0
    \left[
        \frac{\Delta\hat H_l^{\mathrm{sym}}}{\hat Q_\infty}
        +
        \hat H_l^{\mathrm{sym}}
        \left(
            \frac{1}{\hat Q_\infty}
            -
            \frac{1}{\hat Q_l}
        \right)
    \right].
\end{equation}
The first term is controlled by the same harmonic-tail scaling as
Eq.~\eqref{DeltaHTailBoundFinal}. It remains to control the second term. Analogously to the full frequency axis, both $\hat Q_\infty$ and $\hat Q_l$ are
bounded away from zero on the finite interval $\omega\in[0,a]$. Moreover,
\begin{align}
    \hat Q_\infty - \hat Q_l
    &=
    \omega
    \left[
        2-\omega \hat H_l^+(\omega,\bm{k})
    \right]
    \Delta \hat H_l^+(-\omega,\bm{k})
    \notag \\
    &\quad
    -
    \omega
    \left[
        2+\omega \hat H_l^+(-\omega,\bm{k})
    \right]
    \Delta \hat H_l^+(\omega,\bm{k})
    \notag \\
    &\quad
    -
    \omega^2
    \Delta \hat H_l^+(\omega,\bm{k})
    \Delta \hat H_l^+(-\omega,\bm{k}) .
\end{align}
Thus, $\hat Q_\infty-\hat Q_l$ is controlled by the neglected harmonic tail,
with a quadratic remainder that is of higher order in $\Delta \hat H_l^+$.
Consequently, the second contribution in Eq.~\eqref{numericalError} satisfies
\begin{equation}
    \label{densityDifferenceBound}
    \int_0^a
    d\omega\,
    \left|
        \hat n_{1,\infty}^{\mathrm{sym}}(\omega,\bm{k})
        -
        \hat n_{1,l}^{\mathrm{sym}}(\omega,\bm{k})
    \right|
    \leq
    c_4\eta_l,
\end{equation}
where $c_4$ is a computable finite constant depending on the retained frequency
interval and on the distance of the denominators from zero. Combining the preceding estimates gives 
\begin{align}
    \label{totalErrorBound}
    \epsilon_{l,a}(t)
    &\leq
    \frac{c_1}{\pi} R_l(a)
    +
    \left(
        4c_1+\frac{c_4}{\pi}
    \right)\eta_l .
\end{align}
The error bound~\eqref{totalErrorBound} also allows us to reuse the
quadrature-error argument from our previous work. Since the symmetrized
integrand is analytic in a strip around the real axis, whose width is
determined by the Landau pole closest to the real axis, the frequency integral
can be evaluated efficiently using the trapezoidal rule on an equidistant grid
with spacing $h$. Within this strip of analyticity, the trapezoidal rule
converges spectrally~\cite{trefethen:2014}. More precisely, if the integrand is
analytic for $|\operatorname{Im}\omega|<c$, then
\begin{equation}
    |E_h|
    \leq
    \frac{2M}{e^{2\pi c/h}-1},
    \qquad
    M
    \coloneq
    \sup_{|s|<c}
    \int_{-\infty}^{\infty}
    \left|
        \hat n_1^{\mathrm{sym}}(\omega+is,\bm{k})
    \right|
    d\omega .
\end{equation}
Thus, for sufficiently small $h$, the quadrature error can be reduced below the
frequency- and harmonic-truncation errors. In practice, the latter then
dominate the numerical uncertainty.

\subsection{Benchmarking BSL6D}

We finally compare the semi-analytical time-domain solution, obtained by
Fourier inversion of the symmetrized density spectrum
(Fig.~\ref{fig:nRespSym}), with a kinetic BSL6D~\cite{schild2024} simulation.
The simulation is performed on a reduced 2D3V phase-space grid with
$64\times64\times33\times33\times33$ grid points and domain lengths
$2\pi\times40\pi\times8\times8\times8$. We initialize a small-amplitude random
density perturbation with amplitude $\alpha=10^{-5}$ and extract the Fourier
mode with $k_\perp=1$ and $k_z=0.15$, using the same physical parameters as
above. For code verification, we compare the extracted BSL6D signal with the
converged semi-analytical reference solution obtained from the symmetrized
Fourier inversion using the frequency interval $\omega\in[0,15]$ and harmonic
cutoff $l=12$.

\begin{figure}[htbp]
    \centering
    \begin{subfigure}[t]{0.75\columnwidth}
        \centering
        \includegraphics[width=\textwidth]{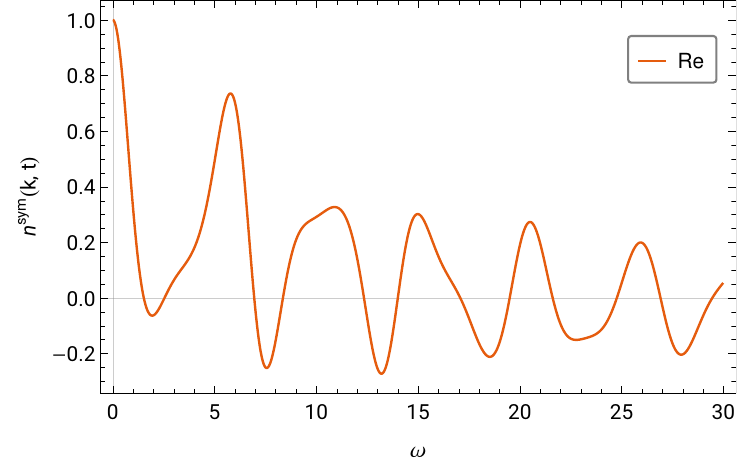}
        \caption{Simulated BSL6D time-domain density response for an initial
        density perturbation with $k_\perp=1$, $k_z=0.15$, and initial
        perturbation amplitude $\alpha=10^{-5}$. The displayed Fourier mode is
        normalized to unity at $t=0$.}
        \label{fig:bsl6d_sim}
        \vspace{0.25cm}
    \end{subfigure}
    \begin{subfigure}[t]{0.75\columnwidth}
        \centering
        \includegraphics[width=\textwidth]{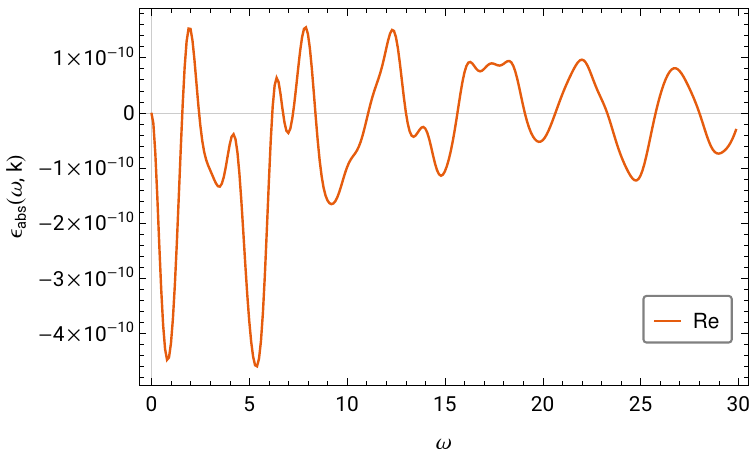}
        \caption{Absolute error of the BSL6D density response in
        Fig.~\ref{fig:bsl6d_sim} relative to the converged semi-analytical
        reference solution on the time interval $t\in[0,30]$.}
        \label{fig:bsl6d_comp}
    \end{subfigure}
    \caption{Time-domain verification of BSL6D against the semi-analytical
    finite-$k_z$ Bernstein density-response benchmark.}
    \label{fig:numerical_comp}
\end{figure}

\section{Limitations of the pure-density benchmark for $k_z=0$}

In the standard Bernstein-wave dispersion-relation test mentioned in the
introduction, strictly perpendicular wave propagation ($k_z=0$) is assumed. For the present time-domain construction, however, this limit is singular: the poles of the self-consistent density response lie on the real frequency axis, as follows from Eqs.~\eqref{nResp} and~\eqref{HkzZero}.

We first note that the nonzero cyclotron singularities of $\hat H_\infty^+$ are
not poles of the full density response. To see this, consider a neighborhood of
the cyclotron harmonic $m\Omega$. The singular term in $\hat H_\infty^+$ can be
separated as
\begin{equation}
    \label{HnearHarmonicm}
    \hat H_\infty^+(\omega)
    =
    \frac{\Gamma_m}{\omega-m\Omega}
    +
    \hat R_m^+(\omega),
\end{equation}
where
\begin{equation}
    \hat R_m^+(\omega)
    \coloneq
    \sum_{p\neq m}
    \frac{\Gamma_p}{\omega-p\Omega}.
\end{equation}
Substituting Eq.~\eqref{HnearHarmonicm} into Eq.~\eqref{nResp} gives
\begin{equation}
    \hat n_1^+(\bm{k},\omega)
    =
    \frac{
        i n_0\left(
        \frac{\Gamma_m}{\omega-m\Omega}
        +
        \hat R_m^+(\omega)
        \right)
    }{
        2-\omega
        \left(
        \frac{\Gamma_m}{\omega-m\Omega}
        +
        \hat R_m^+(\omega)
        \right)
    } .
\end{equation}
Equivalently,
\begin{equation}
    \hat n_1^+(\bm{k},\omega)
    =
    \frac{
        i n_0\left[
        \Gamma_m
        +
        (\omega-m\Omega)\hat R_m^+(\omega)
        \right]
    }{
        (\omega-m\Omega)
        \left[
        2-\omega \hat R_m^+(\omega)
        \right]
        -
        \omega\Gamma_m
    } .
\end{equation}
Thus, for $m\Omega\neq0$ and $\Gamma_m\neq0$,
\begin{equation}
    \lim_{\omega\to m\Omega}
    \hat n_1^+(\bm{k},\omega)
    =
    -\frac{in_0}{m\Omega}.
\end{equation}
The nonzero cyclotron singularities of $\hat H_\infty^+$ are therefore
removable singularities of the full density response. However, near $\omega=0$,
\begin{equation}
    \hat H_\infty^+(\omega)
    =
    \frac{\Gamma_0}{\omega}
    +
    \hat R_0^+(\omega),
\end{equation}
and hence
\begin{equation}
    \hat n_1^+(\bm{k},\omega)
    \sim
    \frac{
        i n_0 \Gamma_0
    }{
        \left(2-\Gamma_0\right)\omega
    },
    \qquad
    \omega\to0 .
\end{equation}
Thus the pure-density initial perturbation contains a genuine zero-frequency
singular contribution in the strictly perpendicular limit. Apart from this zero-frequency contribution, the Bernstein poles $\omega_j$ are determined by the zeros of the Bernstein dispersion function
\begin{equation}
    D(\omega)
    \coloneq
    2-\omega \hat H_\infty^+(\omega),
    \qquad
    D(\omega_j)=0.
\end{equation}
Assuming simple roots, a Taylor expansion near a pole $\omega_j$ gives
\begin{equation}
    \hat n_1^+(\bm{k},\omega)
    \sim
    \frac{A_j}{\omega-\omega_j+i\epsilon},
    \qquad
    A_j
    =
    \frac{i n_0 \hat H_\infty^+(\omega_j)}{D'(\omega_j)} ,
\end{equation}
where
\begin{equation}
    D'(\omega)
    =
    \sum_{p=-\infty}^{\infty}
    \Gamma_p
    \left[
        \frac{\omega}{(\omega-p\Omega)^2}
        -
        \frac{1}{\omega-p\Omega}
    \right].
\end{equation}
The distributional real-frequency limit follows from the Sokhotski-Plemelj formula,
\begin{equation}
    \lim_{\epsilon \to 0^+}
    \frac{1}{\omega-\omega_j+i\epsilon}
    =
    \operatorname{p.v.}\frac{1}{\omega-\omega_j}
    -
    i\pi\delta(\omega-\omega_j).
\end{equation}
Consequently, the causal real-axis response consists of a principal-value part
and singular contributions from the real poles,
\begin{align}
    \lim_{\epsilon\to0^+}
    \hat n_1^+(\omega+i\epsilon,\bm{k})
    =&
    \operatorname{p.v.}
    \left[
    \frac{i n_0 \hat H_\infty^+(\omega)}
    {2-\omega \hat H_\infty^+(\omega)}
    \right]
    -
    i\pi A_0\delta(\omega) \notag \\
    &-
    i\pi
    \sum_j
    A_j\delta(\omega-\omega_j),
\end{align}
where
\begin{equation}
    A_0
    \coloneq
    \frac{i n_0\Gamma_0}{2-\Gamma_0},
\end{equation}
and the sum runs over the nonzero simple Bernstein roots of $D(\omega)$. Using the symmetrization prescription
\begin{equation}
    \hat n_{1}^{\mathrm{sym}}(\omega,\bm{k})
    =
    \lim_{\epsilon\to0^+}
    \left[
        \hat n_1^+(\omega+i\epsilon,\bm{k})
        +
        \hat n_1^+(-\omega+i\epsilon,\bm{k})
    \right],
    \quad
    \omega\in\mathbb R,
\end{equation}
we find
\begin{align}
    \hat n_{1}^{\mathrm{sym}}(\omega,\bm{k})
    &=
    \operatorname{p.v.}
    \left[
    \frac{
        2 i n_0
        \hat H_\infty^{\mathrm{sym}}(\omega,\bm{k})
    }{
        \left(
            2-\omega \hat H_\infty^+(\omega,\bm{k})
        \right)
        \left(
            2+\omega \hat H_\infty^+(-\omega,\bm{k})
        \right)
    }
    \right]
    \notag \\
    &\quad
    -
    i\pi
    \sum_j
    A_j
    \left[
        \delta(\omega-\omega_j)
        +
        \delta(\omega+\omega_j)
    \right].
\end{align}
Thus, after separating the singular pole contributions, the remaining
principal-value spectrum can, in principle, be integrated numerically by
subtracting the real-axis poles from the principle value integrals, as described in our previous work~\cite{isw}.However, in contrast to the finite-$k_z$ case, symmetrization no longerproduces a rapidly decaying smooth spectrum. Instead, the dominant residual tail originates from the $p=0$ contribution and decays only  asymptotically as $\omega^{-2}$. In practice, this renders the strictly perpendicular pure-density benchmark inefficient for high-accuracy time-domain verification.

This observation reflects a compatibility problem between the prescribed pure-density initial perturbation and the undamped $k_z=0$ Bernstein dynamics. A pure density perturbation fixes the initial density, but it does not by itself define a two-sided smooth history generated by a compatible distribution-function perturbation. Thus, the response-function framework does not fail for $k_z=0$ in general; rather, the pure-density perturbation considered here is not an efficient manufactured benchmark in the strictly perpendicular limit.

\section{Electromagnetic generalization}
\label{sec:electromagnetic generalization}

We finally sketch how the present construction extends to electromagnetic
perturbations in homogeneously magnetized plasmas. The resulting formulation is
closely related to the standard treatment in Ref.~\cite{stix}. For
definiteness, we consider a quasineutral electromagnetic plasma model in which
the displacement current has been eliminated,
\begin{align}
    \nabla \cdot \bm B_1 &= 0, \\
    \label{faraday}
    \nabla \times \bm E_1 &= -\partial_t \bm B_1, \\
    \label{maxwell}
    \nabla \times \bm B_1 &= \mu_0 \bm j_1 .
\end{align}
Taking the divergence of Eq.~\eqref{maxwell} gives
$\nabla\cdot\bm j_1=0$, and together with the continuity equation, qasineutrality follows. For each species $s$, the linearized Vlasov equation reads
\begin{equation}
    \label{vlasovMaxwell}
    \partial_t f_{1s}
    +
    \bm v\cdot\nabla f_{1s}
    +
    \frac{q_s}{m_s}
    \left(
        \bm v\times\bm B_0
    \right)\cdot\nabla_{\bm v} f_{1s}
    =
    R_s ,
\end{equation}
with electromagnetic source term
\begin{equation}
    \label{sourceDef}
    R_s
    =
    \frac{q_s}{m_s}
    \left(
        \bm E_1
        +
        \bm v\times\bm B_1
    \right)
    \cdot
    \nabla_{\bm v} f_{0s}.
\end{equation}
For a single spatial Fourier mode, Faraday's law allows us to eliminate $\bm B_1$ in favor of $\bm E_1$. With the same Fourier convention as above, one obtains
\begin{equation}
    \bm B_1^+
    =
    \frac{\bm k\times \bm E_1^+}{\omega},
\end{equation}
and the source can be written as
\begin{equation}
    R_s^+
    =
    \bm M_s^+\cdot\bm E_1^+,
\end{equation}
where
\begin{equation}
    \label{MsDef}
    \bm M_s^+
    \coloneq
    \frac{q_s}{m_s}
    \left[
        \nabla_{\bm v} f_{0s}
        +
        \frac{1}{\omega}
        \left(
            \bm v\times\bm k
        \right)
        \cdot
        \nabla_{\bm v} f_{0s}
    \right].
\end{equation}
The left-hand side of Eq.~\eqref{vlasovMaxwell} has the same characteristic
structure as in the electrostatic problem. Hence, the causal propagator
$\hat G_s^+$ derived above can be used without modification, giving the
Fourier-space response
\begin{equation}
    \label{emf1sol}
    \hat f_{1s}^+
    =
    \hat G_s^+ f_{1s}(0,\bm k,\bm v)
    +
    \hat G_s^+
    \bm M_s^+\cdot\bm E_1^+ .
\end{equation}
In contrast to the electrostatic adiabatic-electron setting considered above,
the field response must now be closed with Maxwell's equations. In Fourier
space, Eqs.~\eqref{faraday} and~\eqref{maxwell} yield
\begin{equation}
    \label{EMwaveEq}
    \left(
        \bm k\bm k
        -
        k^2\bm I
    \right)
    \bm E_1^+
    =
    -i\mu_0\omega\,\bm j_1^+ ,
\end{equation}
where
\begin{equation}
    \bm j_1^+
    =
    \sum_s q_s
    \int d^3v\,\bm v\, \hat f_{1s}^+ .
\end{equation}
Substituting Eq.~\eqref{emf1sol} into the current gives
\begin{equation}
    \bm j_1^+
    =
    \bm j_{\mathrm{init}}^+
    +
    \bm \sigma^+\bm E_1^+ ,
\end{equation}
with initial-condition current
\begin{equation}
    \bm j_{\mathrm{init}}^+
    \coloneq
    \sum_s q_s
    \int d^3v\,
    \bm v\,
    \hat G_s^+
    f_{1s}(0,\bm k,\bm v),
\end{equation}
and kinetic conductivity tensor
\begin{equation}
    \bm \sigma^+
    \coloneq
    \sum_s q_s
    \int d^3v\,
    \bm v\,
    \hat G_s^+
    \bm M_s^+ .
\end{equation}
Here, $\bm v\bm M_s^+$ denotes a dyadic product. The electromagnetic response
is therefore determined by the closed linear system
\begin{equation}
    \label{EMclosedSystem}
    \left[
        \bm k\bm k
        -
        k^2\bm I
        +
        i\mu_0\omega\,\bm \sigma^+
    \right]
    \bm E_1^+
    =
    -i\mu_0\omega\,\bm j_{\mathrm{init}}^+ .
\end{equation}
Solving Eq.~\eqref{EMclosedSystem} amounts to inverting a field-response matrix.
The resulting $\bm E_1^+$ can then be inserted into Eq.~\eqref{emf1sol}, and
the time-domain distribution function is obtained by the same Fourier-inversion
procedure as in the electrostatic case.

\section{Conclusion}

In this work, we have extended our semi-analytical framework for computing
distribution-function and moment responses~\cite{isw} to uniformly magnetized
plasmas. The resulting formulation enables time-domain benchmarks for kinetic
solvers that probe dynamics both parallel and perpendicular to the background
magnetic field.

The main result is an explicit semi-analytical benchmark for the quasineutral
Bernstein-wave density response at finite $k_z$. This regime is damped by
parallel phase mixing and is therefore not covered by the standard undamped
$k_z=0$ Bernstein-wave dispersion-relation tests. The extension is non-trivial, as the magnetized response involves helical characteristics, cyclotron-harmonic
coupling, and a more delicate real-frequency pole structure. We derived a
corresponding error estimate that controls both the cyclotron-harmonic
truncation error and the frequency-domain truncation error. As a consistency
check, the unmagnetized result is recovered in the limit $B_0\to0$ for
$k_z\neq0$.

We also showed that the strictly perpendicular limit $k_z=0$ requires a
separate interpretation. In this undamped case, the symmetrized
principal-value contribution decays only algebraically in Fourier space and is
therefore not well suited for efficient direct numerical Fourier inversion.
For the pure-density perturbation considered here, this reflects a
compatibility issue between the prescribed initial data and the undamped
Bernstein dynamics, rather than a failure of the response-function framework
itself.

Finally, we outlined the extension to electromagnetic perturbations. This
generalization is conceptually straightforward but algebraically even more involved: the same magnetized propagator determines the kinetic response, while
Maxwell's equations close the system through a field-response matrix. The
framework therefore provides a controlled route to time-domain benchmarks for
kinetic solvers in both magnetized electrostatic and electromagnetic plasma
regimes.

\section{Acknowledgments}
    This work has been carried out partly within the framework of the EUROfusion
    Consortium, funded by the European Union via the Euratom Research and Training
    Programme (Grant Agreement No 101052200 – EUROfusion). Support has also been
    received by the EUROfusion High Performance Computer (Marconi-Fusion). Views and
    opinions expressed are however those of the author(s) only and do not
    necessarily reflect those of the European Union or the European Commission.
    Neither the European Union nor the European Commission can be held responsible
    for them.  Numerical simulations were performed at the MARCONI-Fusion
    supercomputer at CINECA, Italy, and at the HPC system at the Max Planck
    Computing and Data Facility (MPCDF), Germany.

\footnotesize
\bibliographystyle{elsarticle-num}
\bibliography{references}

\end{document}